\documentstyle[11pt,newpasp,twoside,epsf,rotate]{article}
\def\edcomment#1{\iffalse\marginpar{\raggedright\sl#1\/}\else\relax\fi}
\reversemarginpar

\begin{document}
\title{Population synthesis of old neutron stars in the Galaxy}
 \author{S.B. Popov}
\affil{Sternberg Astronomical Institute, Universiteskii Pr. 13,
119899, Moscow, Russia; e--mail: polar@xray.sai.msu.su}
\author{M. Colpi}
\affil{Dept. of Physics, Univ. of Milan, Via Celoria
16, 20133 Milan, Italy;  e--mail: monica@pccolpi.uni.mi.astro.it}
\author{A. Treves}
\affil{Dipartimento di Scienze, Univ.  dell'Insubria,
Via Lucini 3, 22100, Como, Italy; e--mail: treves@mi.infn.it}
\author{R. Turolla}
\affil{Dept. of Physics, Univ. of Padova, Via Marzolo 8,
35131 Padova, Italy}
\author{V.M. Lipunov}
\affil{Dept. of Physics, Moscow State Univ.; Sternberg Astronomical
Institute, Universiteskii Pr. 13,
119899, Moscow, Russia}
\author{M.E. Prokhorov}
\affil{Sternberg Astronomical Institute, Universiteskii Pr. 13,
119899, Moscow, Russia; e--mail: mystery@sai.msu.su}

\begin{abstract}
The paucity of  old isolated accreting neutron stars in
ROSAT observations is used to derive a lower limit on the mean
velocity of neutron stars at birth. The secular evolution of the
population is simulated following the paths of a statistical
sample of stars for different values of the initial kick velocity,
drawn from an isotropic Gaussian distribution with 
mean  velocity
$0\leq \langle V\rangle\leq 550$ ${\rm km\,s^{-1}}$. The
spin--down, induced by dipole losses and the interaction with
the ambient medium, is tracked together with the dynamical
evolution in the Galactic potential, allowing for the
determination of the fraction of stars which are, at present, in
each of the four possible stages: Ejector, Propeller, Accretor,
and Georotator. Taking  from the ROSAT All Sky Survey an upper
limit of $\sim 10$ accreting neutron stars within $\sim 140$ pc from the
Sun, we infer a lower bound for the mean  kick velocity, 
$ \langle V\rangle\ga 200-300$ ${\rm km\,s^{-1}}$.
The same conclusion is reached for both a constant 
($B\sim 10^{12}$ G) and a magnetic field decaying exponentially with a
timescale $\sim 10^9$ yr. 
Present results,
moreover, constrain the fraction of low velocity stars, which
could have escaped pulsar statistics, to $\la 1\%$.
\end{abstract}

\section{Introduction}

Isolated neutron stars (NSs) are expected to be as many as
$10^8$--$10^9$, a non--negligible fraction  of the
total stellar content of the Galaxy.
The number of observed radio
pulsars is now $\sim$ 1,000. Since the pulsar lifetime is $\sim
10^7$ yr, this implies that the bulk of the NS population, mainly
formed of old objects, remains undetected as yet. Despite
intensive searches at all wavelengths, only a few (putative)
isolated NSs which are not radio pulsars 
(or soft $\gamma$ repeaters)
have been recently
discovered in the X--rays with ROSAT 
(Walter, Wolk \& Neuhauser
1996; Haberl et al. 1998; Neuhauser \& Trumper 1999). 
The extreme X--ray to optical flux ratio ($> 10^3$)
makes the NS option rather robust, but the exact nature of their
emission is still controversial. Up to now, two main possibilities
have been suggested, either relatively young NSs
radiating away their residual internal energy or much aged NSs
accreting the interstellar medium (ISM). Both options have advantages and
drawbacks. Standard cooling atmosphere models fail to predict in a natural
way the spectrum  of the best studied object, RX
J1856-3754 (see Walter et al. this volume).
Accretion models require instead  a very low
NS velocity relative to the ISM
($v < 20 $ km\,s$^{-1}$)  in order  to produce the luminosities  
inferred from ROSAT data
(see Walter et al. this volume). We  feel
that a more thorough analysis of the statistical properties of NSs
can be useful in providing indirect evidence in
favor or against the accretion scenario.

As discussed by Lipunov (1992), 
isolated NSs can be classified into four main types:
Ejectors, Propellers, Accretors and Georotators. In Ejectors the relativistic
outflowing momentum flux is always larger than the ram pressure of the
surrounding  material so they never accrete and are
either active or dead pulsars, still spun down by dipole
losses. In Propellers the
incoming matter can penetrate down to the Alfven radius, $R_A$, but no
further because of the centrifugal barrier, and stationary
inflow can not occur, but the piling up of the material at the
Alfven radius may give rise to (supposedly short) episodes of
accretion
(Treves, Colpi \& Lipunov 1993; Popov 1994). 
Steady accretion is also impossible in Georotators where
(similarly to the Earth) the Alfven radius exceeds the accretion
radius, so that magnetic pressure  dominates everywhere over the
gravitational pull. It is the combination of the star period,
magnetic field and velocity that decides  which type a given
isolated NS belongs to
and, since both $P$, $B$ and $V$ change during the star evolution,
a NS can
go through different stages in its lifetime. 

While the dynamical evolution of NSs in the Galactic potential was studied by
several authors (sse e.g. Madau \& Blaes 1994; Zane et al. 1995), 
little attention was
paid to the NSs magneto--rotational evolution. Recently, this issue
was discussed in some detail by 
Livio, Xu \& Frank (1998) and Colpi et al. (1998).
Goal of this investigation is to consider these two issues
simultaneously, coupling the dynamical and the magneto--rotational
evolution for the isolated NS population.

The possibility that the
low--velocity tail is underpopulated with respect to what was
previously assumed should be seriously taken into account. It is
our aim to revise the estimates on the number of old accreting
neutron stars in the Galaxy
in the light of these new data, in the attempt to
reconcile theoretical predictions with present ROSAT limits
(Ne\"uhauser \& Tr\"umper 1999). 

\section{The Model}

In this section we summarize  the main hypothesis
introduced to track the evolution of single stars and describe
shortly the technique used
to explore their statistical properties, referring to 
Popov \& Prokhorov (1998) 
for details on spatial evolution calculations
and to Konenkov \& Popov (1997) 
and Lipunov \& Popov (1995) for details of magneto-rotational evolution.

\subsection{Dynamical evolution}

The dynamical evolution of each single star in the  Galactic potential 
(taken in the form proposed by Miyamoto \& Nagai 1975) is
followed
solving its equations of motion. 

The period evolution depends on both the star
velocity and the local density of the interstellar medium, any
attempt to investigate the statistical properties of the NS
population should incorporate a detailed model of the ISM
geography. Unfortunately the distribution of molecular and atomic
hydrogen in the Galaxy is highly inhomogeneous. 
Here we use the analytical distributions from 
Bochkarev (1992) and Zane et al. (1995) for the hydrogen density
$n(R,Z)$.
Within
a region of $\sim 140$ pc around the Sun, the ISM is underdense,
and we take $n=0.07 \ {\rm cm}^{-3}$.

In our model we assume that the NS birthrate is constant in time
and proportional in magnitude to the square of the local gas
density. 

Neutron stars at birth have a circular velocity determined by the
Galactic potential. Superposed to this ordered motion a kick
velocity is imparted in a random direction. 
We use here an isotropic Gaussian distribution (relative to
the local circular speed) 
with dispersion $\sigma_V$, simply as a mean to model
the true pulsar distribution at birth (see e.g. Cordes \& Chernoff 
1998).
The mean velocity 
$\langle V\rangle=(8/\pi)^{1/2}\sigma_V$ 
is varied in the interval 0--550 km\, s$^{-1}$. 

\subsection{Accretion physics and period evolution}

The accretion rate was calculated according to the Bondi formula

\begin{equation}
\dot M=
\frac{2\pi (GM)^2 m_p n(R,Z)}{(V^2+V_s^2)^{3/2}}\simeq 10^{11}\,
n\,
v_{10}^{-3}\ {\rm g\, s}^{-1}
\end{equation}
where $m_p$ is proton mass, the sound speed $V_s$ is always 10
km$\,{\rm s}^{-1}$ and $v_{10} = (V^2+V^2_s)^{1/2}$ in units of 10
km$\,
{\rm s}^{-1}$. $M$ and $R$ denote the NS
mass and radius, which we take equal to $1.4\, M_\odot$ and 10 km,
respectively, for all stars. 

All neutron stars are assumed to be born  with a period
$P(0) =$ 0.02 s, and a magnetic moment either
$\mu_{30}=1$ or $ \mu_{30}=0.5$, where $\mu_{30}=\mu/10^{30} {\rm
G}\, {\rm cm}^3$. 

\begin{figure}[t]
\epsfxsize=0.8\hsize
\centerline{\rotate[r]{\epsfbox{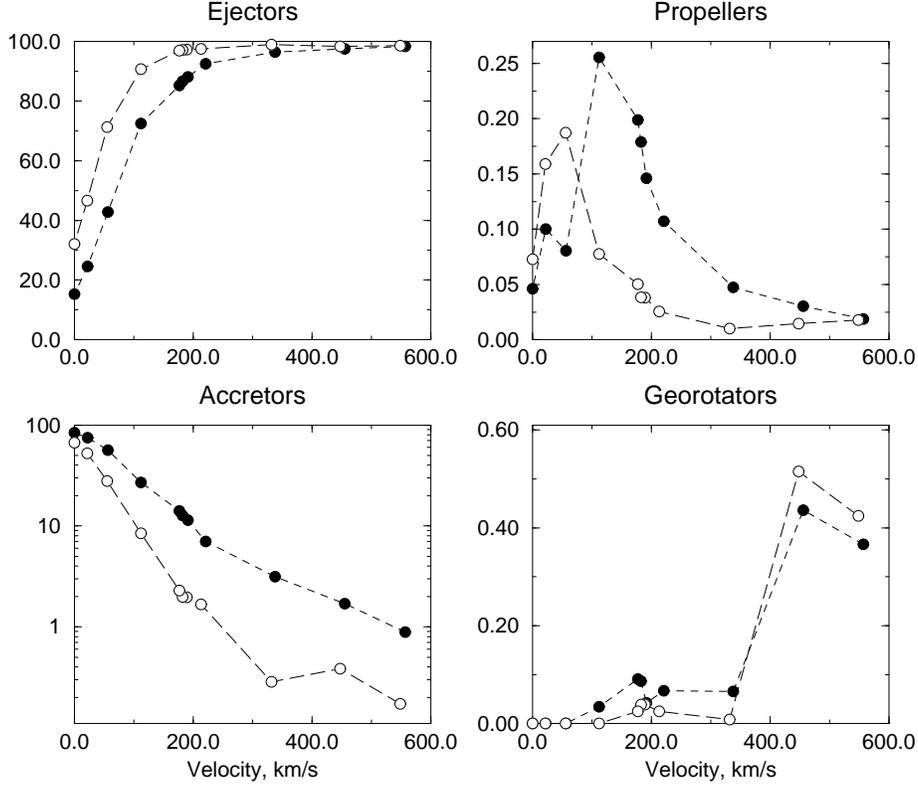}}}
\caption{Fractions of NSs in the different stages vs. the mean kick
velocity for $\mu_{30}=0.5$ (open circles) and $\mu_{30}=1$
(filled circles); typical statistical uncertainty for ejectors and
accretors is $\sim $ 1-2\%.}
\end{figure}

\begin{figure}[t]
\epsfxsize=0.9\hsize
\centerline{\rotate[r]{\epsfbox{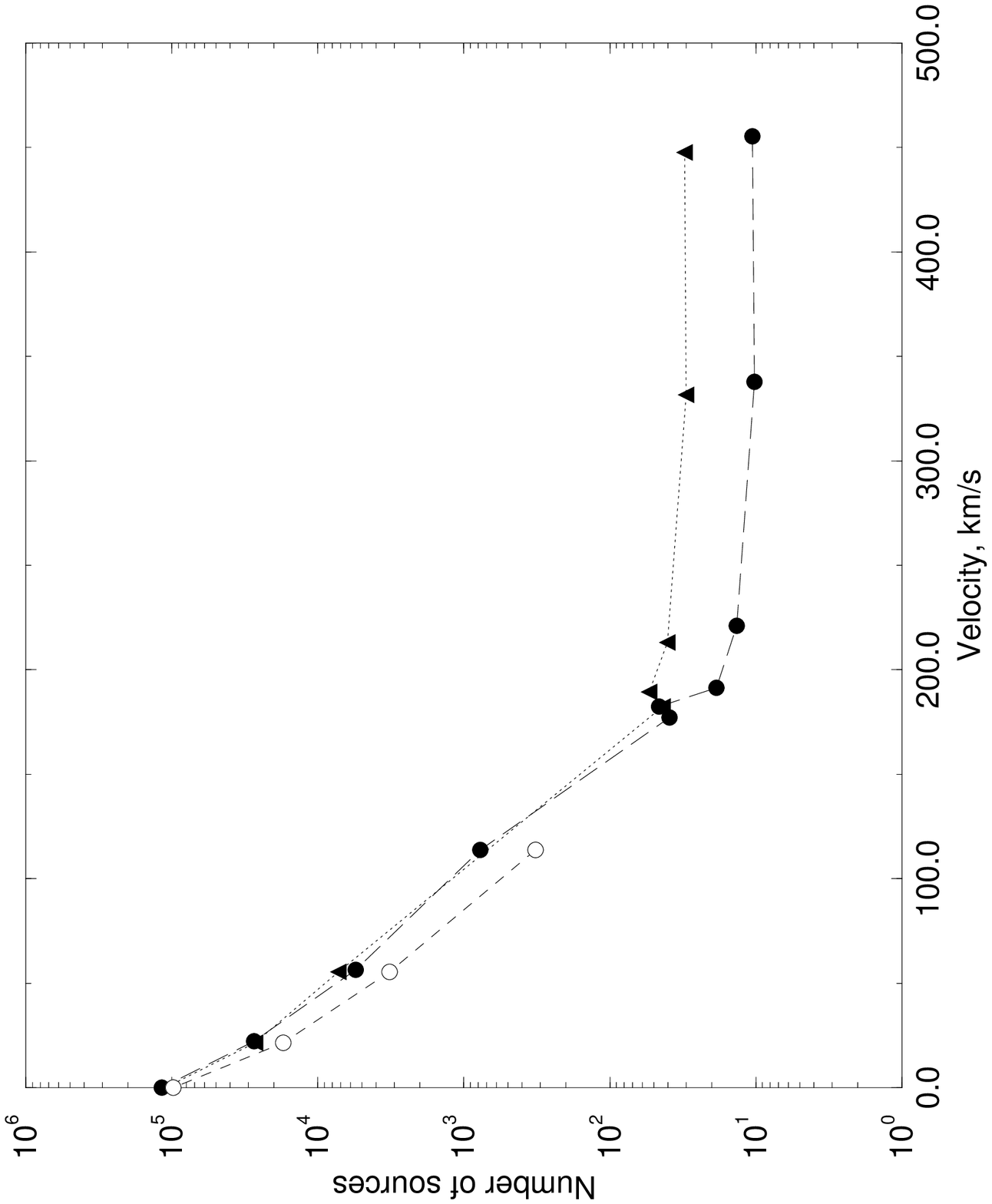}}}
\caption{Number of accreting NSs in the Solar vicinity 
for a constant ($\mu_{30}=1$, filled circles;
$\mu_{30}=0.5$, opened circles) 
and decaying field
($t_d = 2.2\times 10^9 $ yr, triangles).}
\end{figure}

In the ejector phase the energy losses
are due to magnetic dipole radiation.
When the gravitational energy density of the incoming interstellar gas
exceeds
the outward momentum flux  at the accretion radius,
$R_{ac}\simeq 2GM/v^2$, matter starts to fall in. This happens when the period
reaches the critical value
\begin{equation}\label{petop}
P_{E}(E\to P)\simeq 10\, \mu_{30}^{1/2}\, n^{-1/4}\,v_{10}^{1/2}\ \rm {s}.
\end{equation}
When $P>P_{E}(E\to P)$  
the NS is in the propeller phase, rotational energy
is lost and the period keeps increasing at a rate taken from
Shakura (1975).

As the star moves through the inhomogeneous  ISM  a transition from
the propeller back to the  ejector phase may occur
if the period attains the critical value

\begin{equation}
P_E(P\to E)\simeq 3\, \mu_{30}^{4/5} v_{10}^{6/7} n^{-2/7}
\rm{s}\, .
\end{equation}
Note that the transitions $P\to E$ and $E\to P$ are not
symmetric as first discussed  by Shvartsman in the early '70s.

Accretion onto the star surface occurs when the corotation radius
$R_{co}=(GM\,P^2/4\pi^2)^{1/3}$ becomes larger than the Alfven
radius (and $R_A<R_{ac}$, see below). This implies that braking
torques have  increased the period up to
\begin{equation}
P_{A}(P\to A)\simeq 420\, \mu_{30}^{6/7}\, n^{-3/7}
\, v_{10}^{9/7}\ \rm s\, .
\end{equation}
As soon as the NS enters the accretor phase, torques produced
by stochastic angular momentum exchanges in the ISM
slow down the star
rotation at the equilibrium period
\begin{equation}
 P_{eq}=2.6\times 10^{3}\,
 v^{-2/3}_{(t)10}\,\mu_{30}^{2/3}\, n^{-2/3}\,v_{10}^{13/3}\ {\rm s}
\end{equation}
where $v_{(t)}$ the turbulent velocity of the ISM 
(Lipunov \& Popov 1995; Konenkov \& Popov 1997).

At the
very low accretion rates expected for fast, isolated NSs, it could
be that the Alfven radius is larger than the accretion radius. The
condition $R_A<R_{ac}$ translates into a limit for the star
velocity
\begin{equation}
v <410 \, n^{1/10}\,\mu_{30}^{-1/5} \
{\rm km\, s}^{-1}\, .
\end{equation}

\section{Results and discussion}

\subsection {The NS census for a non--decaying field}

We consider two representative values for the (costant) magnetic dipole 
moment, 
$\mu_{30}= 0.5$ and $\mu_{30}=1$.
The present fraction of NSs in the
Ejector and Accretor stages as a function of the mean kick
velocity is shown  in figure 1. 

\begin{figure}[t]
\epsfxsize=0.8\hsize
\centerline{\rotate[r]{\epsfbox{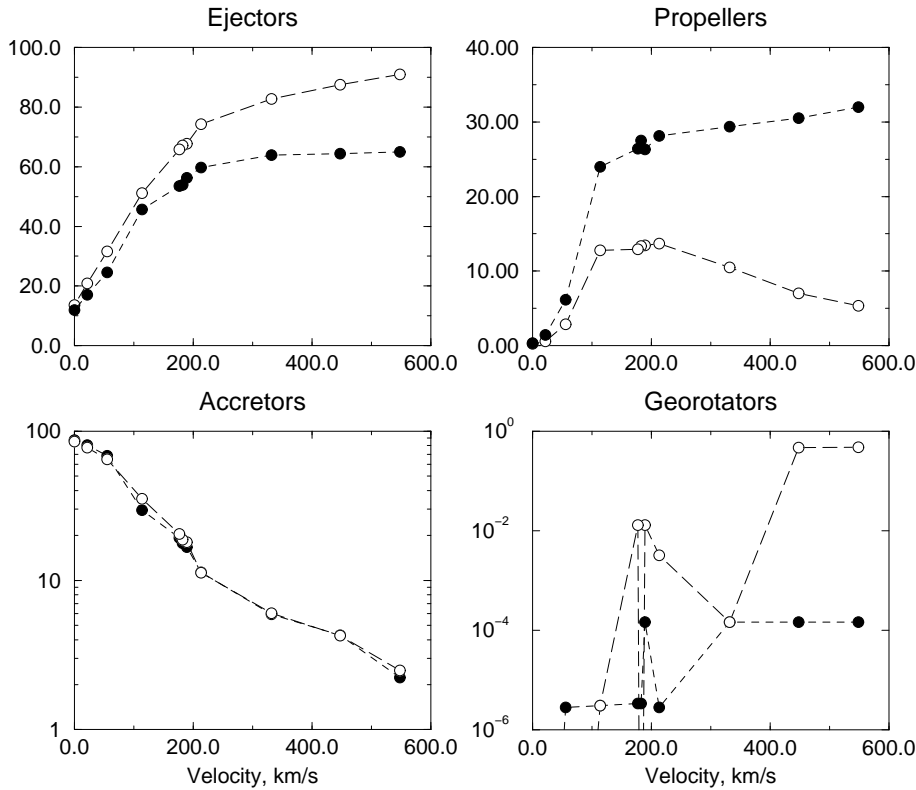}}}
\caption{Fractions of NSs in the different stages vs. the average kick
velocity for a decaying field with an e--folding time
$t_d=2.2\times 10^9$ yrs (open circles) and $t_d=1.1\times 10^9$ yrs
(filled circles).}
\end{figure}

\begin{figure}[t]
\epsfxsize=0.9\hsize
\centerline{\rotate[r]{\epsfbox{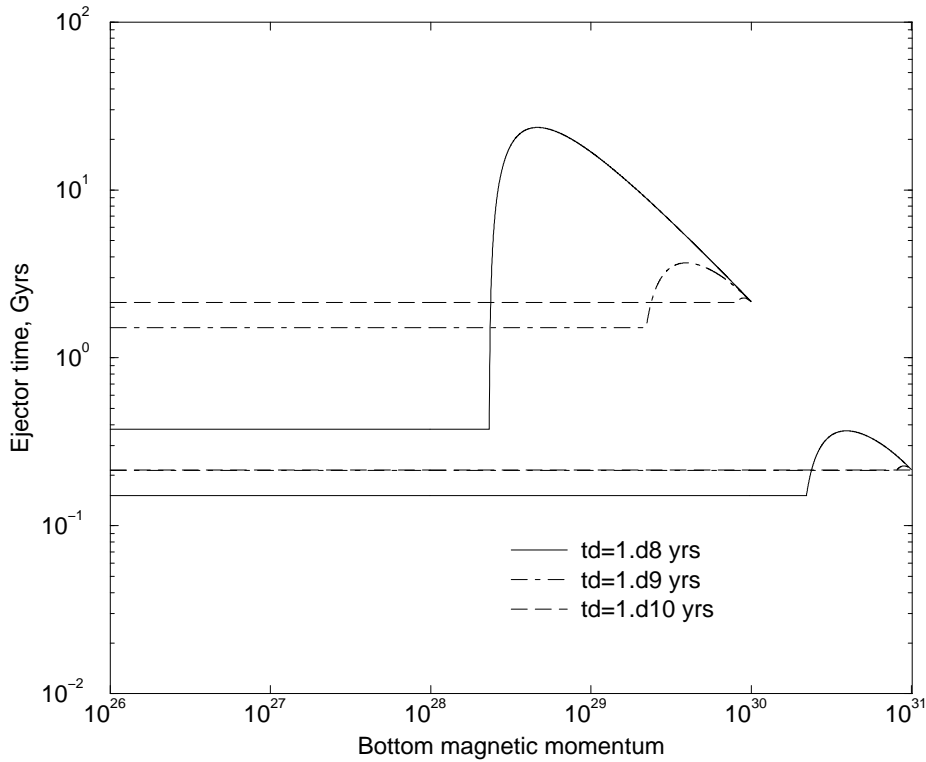}}}
\caption{Ejector time vs. bottom magnetic momentum
for two values of initial magnetic momentum and different decay timescales,
$t_d= 10^8$, $10^9$, and $10^{10}$ yr.}
\end{figure}

Here, and in the following the total number of
Galactic NSs was assumed to be $10^9$. 
A total number $\sim 10^9$ appears to be consistent with the nucleosynthesis
and chemical evolution of the Galaxy, while $10^8$ is derived
from radio pulsars observations. It is 
uncertain if all NSs experience an active radio pulsar phase, due to
low initial magnetic fields or long periods, or to 
the fall--back in the aftermath of the supernova explosion.
There is a serious possibility that the total number of NSs derived
from radio pulsar statistics is only a lower limit.

In order to compare the expected number of accreting ONSs with the
ROSAT All Sky Survey (RASS) results, we evaluated the number of
those ONSs, within 140 pc from the Sun, producing an unabsorbed
flux of $10^{-13}$ erg cm$^{-2}$ s$^{-1}$ or higher at energies $\sim 100$
eV.
The results are illustrated in figure 2.
The main point is that for mean
velocities below 200 km s$^{-1}$ the number of  ONSs with a flux
above the RASS detection limit would exceed 10. Most recent
analysis on the number of isolated NSs in the RASS 
(Ne\"uhauser \& Tr\"umper 1999) 
indicate that the upper limit is below 10. 

An important aspect is that our results exclude
the possible presence of a consistent low--velocity population at
birth, which exceeds that contained in the 
gaussian with 
$\langle V\rangle> 200$ km s$^{-1}.$ 

\subsection {The NS census for a decaying field}

We refer here only
to a very simplified picture of the field decay
in which $B(t) = B(0)\exp{(-t/t_d)}$.
Calculations have been performed for $t_d=1.1\times 10^9$
yr, $t_d=2.2\times 10^9$ yr and $\mu_{30}(0) =1$. 
Results are shown in figure 3. 

For some values of $t_d$ and bottom field most of NS
can stay at the Ejector stage, and the number of accretors and Propellers
would not be increased. 
We show this analytical estimates graphically in figure 4,
where the Ejector time, $T_E$, is plotted vs. bottom magnetic momentum
for constant velocity and ISM density
($n=1$ ${\rm cm}^{-3}$, $v=10$ km/s), different $t_d$
and two values of the initial magnetic momentum, $10^{30}$ and $10^{31}$
G cm$^3$ (see Popov \& Prokhorov 1999).

Summarizing, we can conclude that, although both the initial distribution
and the subsequent evolution of the
magnetic field strongly influences the NS census and should be accounted
for, the lower bound on the 
average kick derived from ROSAT surveys is not
very sensitive to $B$, at least for not too extreme values of $t_d$ and
$\mu(0)$, within this model. 

\section{Conclusions}

In this paper we have investigated how the present distribution of
neutron stars in the different stages (Ejector, Propeller,
Accretor and Georotator) depends on the star mean velocity at
birth. On the basis of a total of $\sim 10^9$ NSs, the fraction of
Accretors  was used to estimate the number of sources within 140
pc from the Sun which should have been detected by ROSAT. Most
recent analysis of ROSAT data indicate that no more than $\sim 10$
non--optically identified sources can be accreting ONSs. This
implies that the 
average velocity of the NS population at birth has to
exceed $\sim 200 \ {\rm km\, s^{-1}}$, a figure which is
consistent with those derived from radio pulsars statistics. We
have found that this lower limit on the mean kick velocity is
substantially the same either for a constant or a decaying
$B$--field, unless the decay timescale is shorter than $\sim 10^9$
yr. Since observable accretion--powered ONSs are slow objects, our
results exclude also the possibility that the present velocity
distribution of NSs is richer in low--velocity objects with
respect to a Maxwellian. The paucity of accreting ONSs seem
therefore to lend further support in favor of neutron stars as
very fast objects.

\acknowledgments
Work partially supported by the European
Commission under contract ERBFMRX-CT98-0195.
The work of S.P., V.L and M.P. was supported by grants
RFBR 98-02-16801 and INTAS 96-0315.
S.P. and V.L. gratefully acknowledge the University of Milan and
of Insubria (Como) for support during their visits. 
S.P. also acknowledge organizers of the IAU 195.


\begin{references}
\reference
Bochkarev, N.G. 1992, Basics of the ISM Physics (Moscow University Press)
\reference
Colpi, M., Turolla, R., Zane, S., \& Treves, A. 1998,
ApJ 501, 252
\reference
Cordes, J.M.,\& Chernoff, D.F.  1998, ApJ 505, 315
\reference
Haberl, F., Motch, C., \& Pietsch, W. 1998, Astron.
Nachr. 319, 97
\reference
Konenkov, D.Yu., \& Popov, S.B. 1997, PAZh, 23, 569
\reference
Lipunov, V.M. 1992, Astrophysics of Neutron Stars (Springer \& Verlag)
\reference
Lipunov, V.M., \& Popov, S.B. 1995, AZh, 71, 711
\reference
Livio, M., Xu, C., \& Frank, J. 1998 ApJ, 492, 298
\reference
Madau, P., \& Blaes, O. 1994, ApJ, 423, 748
\reference
Miyamoto, M., \& Nagai, R. 1975, Pub. Astr. Soc. Japan, 27, 533
\reference
Ne\"uhauser, R., \& Tr\"umper, J.E. 1999, A\&A, 343, 151
\reference
Popov, S.B. 1994, Astron. Circ., N1556, 1
\reference
Popov, S.B., \& Prokhorov, M.E. 1998, A\&A, 331, 535
\reference
Popov, S.B., \& Prokhorov, M.E. 1999, astro-ph/9908212
\reference
Shakura, N.I. 1975, PAZh, 1, 23
\reference
Treves, A., Colpi, M., \& Lipunov, V.M. 1993, A\&A, 269, 319
\reference
Walter, F.M., Wolk, S.J., \& Ne\"uhauser, R. 1996, Nature,
379, 233
\reference
Zane, S., Turolla, R.,  Zampieri, L.,
Colpi, M., \& Treves, A. 1995, ApJ,  451, 739
\end{references}
\end{document}